\documentclass[stef,twoside]{stefano}
\usepackage{amsmath}
\usepackage{amssymb}
\usepackage{graphics}
\usepackage{rotating}
\usepackage{cite}
\usepackage{color}

\def\makeheadbox{{%
\hbox to0pt{\vbox{\baselineskip=10dd\hrule\hbox
to\hsize{\vrule\kern3pt\vbox{\kern3pt \hbox{  {\sc journal of
optics a} {\bf 10}, 115001-5 (2008).} \hbox{ {\sc
{\color{blue}{dma}}[{\color{black}{imecc}}]{\color{red}{UniCamp}}
} \hspace*{10.4cm} {\color{blue}{$\boldsymbol{\Sigma \delta
\Lambda}$}} }
\kern3pt}\hfil\kern3pt\vrule}\hrule}%
\hss}}}

\def\0{\mbox{\tiny $0$}}
\def\1{\mbox{\tiny $1$}}
\def\2{\mbox{\tiny $2$}}
\def\3{\mbox{\tiny $3$}}
\def\4{\mbox{\tiny $4$}}
\def\5{\mbox{\tiny $5$}}
\def\6{\mbox{\tiny $6$}}
\def\7{\mbox{\tiny $7$}}
\def\8{\mbox{\tiny $8$}}
\def\9{\mbox{\tiny $9$}}
\def\I{\mbox{\tiny $I$}}
\def\II{\mbox{\tiny $II$}}
\def\III{\mbox{\tiny $III$}}
\def\L{\mbox{\tiny $L$}}
\begin{document}
%
%%%%%%%%%%%%%%%%%%%%%%%%%%%%%%%% PAPER %%%%%%%%%%%%%%%%%%%%%%%%%%%%%%%%%%%%%

\title{\Large LOCALIZED BEAMS AND DIELECTRIC BARRIERS}
%\subtitle{}

\author{
Stefano De Leo\inst{1}
%\thanks{Partially supported by the FAPESP grant 99/09008--5.}
%\and Antonio Moro\inst{2}
%\thanks{Supported by a CAPES PhD fellowship.}
%\and
%Celso C. Nishi\inst{2}
\and Pietro P. Rotelli\inst{2} }

\institute{
Department of Applied Mathematics, State University of Campinas\\
PO Box 6065, SP 13083-970, Campinas, Brazil\\
{\em deleo@ime.unicamp.br}
%{\em ducati@ime.unicamp.br}
%\and
%School of Mathematics, University of Loughbourogh\\
%Leicestershire, LE11 3TU, UK\\
%{\em a.moro@lboro.ac.uk}
%\and Department of Cosmic Rays and Chronology, State University of
%Campinas\\
%PO Box 6165, SP 13083-970, Campinas, Brazil\\
%{\em ccnishi@ifi.unicamp.br}
\and
Department of Physics, INFN, University of Lecce\\
PO Box 193, 73100, Lecce, Italy\\
{\em rotelli@le.infn.it} }

%%%%%%%%%%%%%%%%%%%%%%%%%%%%%%%%%%%%%%%%%%%%%%%%%%%%%%%%%%%%%%%%%%%%%%%%%%%
%%%%%%%%%%%% DATE ABSTRACT PACS % %%%%%%%%%%%%%%%%%%%%%%%%%%%%%%%%%%%%%%%%%

\date{Submitted: {\em May, 2008}. Revised: {\em July, 2008}. }
%- Revised version:  {\em April, 2004} }
% Warning: Where is the date?

\abstract{Recalling the similarities between the Maxwell equations
for a transverse electric wave in a stratified medium and the
quantum mechanical Schr\"odinger equation in a piece-wise
potential, we investigate the analog of the so called particle
limit in quantum mechanics. It is shown that in this limit the
resonance phenomena are lost since individual reflection and
transmission terms no longer overlap. The result is a stationary
zebra-like response with the intensity in each stripe calculable.}

%%%%%%%%%%%%%%%%%%%%%%%%%%%%%%%%%%%%%%%%%%%%%%%%%%%%%%%%%%%%%%%%%%%%%%%
%%%%%%%%%%%%%%%%%%%%%%%%%%%%%%%%%%%%%%%%%%%%%%%%%%%%%%%%%%%%%%%%%%%%%%%

%%%%%%%%%%%%%%%%%%%%%%%%%%%%%%%%%%%%%%%%%%%%%%%%%%%%%%%%%%%%%%%%%%%%%%%
%%%%%%%%%%%%%%%%%%%%%%%%%%%%%%%%%%%%%%%%%%%%%%%%%%%%%%%%%%%%%%%%%%%%%%%


% Warning: No PACS code given

%02.10.Hh Rings and algebras
%02.10.Ud Linear algebra
%02.10.Yn Matrix theory

%02.30.Hq Ordinary differential equations
%02.30.Jr Partial differential equations
%02.30.Tb Operator theory

%03.65.-w Quantum mechanics
%03.65.Ca Formalism
%03.65.Ta Foundations of quantum mechanics;
%03.65.Xp Tunneling, traversal time,quantum Zeno dynamics

%12.15.F Quarks and lepton masses and mixing
%14.60.Pq Neutrino mass and mixing

%42.25.Bs Wave propagation, transmission and absorption
%42.25.Gy Edge and boundary effects; reflection and refraction
%42.50.Xa Optical tests of quantum theory

%\offprints{~Stefano De Leo.}

\titlerunning{\sc localized beams and dielectric barriers}

\maketitle

\section*{I. INTRODUCTION}

There exist many unanswered questions in potential theory quantum
mechanics. Amongst these is the existence of multiple diffusion
phenomena\cite{DEL1,DEL2}, the Hartman effect with its apparent
violation of causality\cite{HAR,REC}, the importance, if any, of
wave packets in oscillation phenomena\cite{DELNEU1,DELNEU2}. Most
of these lack direct experimental measurements. It is therefore
extremely instructive to study that class of Maxwell equations
which are analogous to the Schr\"odinger equation. These analogies
are well known in optics e.g. we often find references to
tunneling and resonance phenomena. However, their relevance to
quantum mechanics has not been fully exploited. This paper studies
an example of the above analogy.

From the Maxwell equations, we can obtain differential equations
which the electric and the magnetic vector must separately
satisfy\cite{BORN}. For example, for the electric field
$\boldsymbol{E}$, in the case of no charges or currents, one has
\begin{equation}
\label{eqE}
\nabla^{^{2}}\boldsymbol{E} -\frac{\epsilon
\mu}{c^{\2}}\,\partial_{tt}\boldsymbol{E}+(\nabla \ln \mu) \times
(\nabla \times \boldsymbol{E}) + \nabla (\boldsymbol{E} \cdot
\nabla \ln \epsilon)  =  0\,\,.
\end{equation}
The corresponding equation for the magnetic field $\boldsymbol{H}$
is obtained by making the changes $\epsilon \leftrightarrow \mu$
and $\boldsymbol{E}\to \boldsymbol{H}$. We shall confine our
attention to the study of a medium characterized by a real (no
attenuation) refractive index whose properties are constant
throughout each plane perpendicular to the chosen direction
$\boldsymbol{z}$, stratified medium\cite{ABE},
\begin{equation}
n(z) = \left\{\,n_{\I}\,\,\,\mbox{for $z<0$}\,\,\,,\,\,\,\,\,
n_{\II}\,\,\,\mbox{for $0<z<L$}\,\,\,,\,\,\,\,\,
n_{\III}=n_{\I}\,\,\,\mbox{for $z>L$}\, \right\}\,\,,
\end{equation}
and for which $\mu$ assumes the same value in all three regions.
By taking the plane of incidence to be the $y$-$z$ plane, for a
monochromatic, $\exp(-i\omega t)$, transverse electric wave
($E_{\2,\3}=0$),  Eq.(\ref{eqE}) reduces to
\begin{equation}
\partial_{yy}E_{\1}(y,z)+\partial_{zz}E_{\1}(y,z)+n^{\2}(z)
k^{\2}E_{\1}(y,z)=0\,\,,
\end{equation}
with $k=\omega/c$ and where we have used the fact  that $\nabla
\cdot (\epsilon \boldsymbol{E})=0$ implies that $E_{\1}$ is a
function of $y$ and $z$ only. The components of the magnetic
vector can be determined by using
 $\nabla \times \boldsymbol{E} = -\,\partial_t\,
(\mu \boldsymbol{H})/c$,
\begin{equation}
\left\{\,H_{\2}(y,z)\,,\,H_{\3}(y,z)\,\right\} = \frac{i}{k\mu}\,
\left\{\,-\,\partial_zE_{\1}(y,z)\,,\,\partial_yE_{\1}(y,z)\,\right\}\,\,.
\end{equation}
The geometry of our problem is schematically represented in the
following picture:

\begin{picture}(300,150) \thinlines
\put(50,45){\vector(0,1){30}} \put(48,77){$H_{\2}$}
\put(50,45){\vector(1,1){20}} \put(71,68){$E_{\1}$}
\put(50,45){\vector(1,0){30}} \put(83,41){$H_{\3}$}
\put(212,10){\vector(0,1){58}} \put(210,76){$y$}
\put(212,10){\vector(1,1){38}} \put(252,52){$x$}
\put(62,10){\vector(1,0){210}} \put(280,8){$z$}
\put(90,120){\mbox{\Large $n_{\mbox{\footnotesize $I$}}$}}
\put(155,122){\mbox{\Large $n_{\mbox{\footnotesize $II$}}$}}
\put(216,120){\mbox{\Large $n_{\mbox{\footnotesize $I$}}$}}
 \put(119,0){$0$} \put(159,0){$L$}  \thicklines
\put(120,10){\line(0,1){90}} \put(125,10){\line(0,1){90}}
\put(130,10){\line(0,1){90}} \put(135,10){\line(0,1){90}}
\put(140,10){\line(0,1){90}} \put(145,10){\line(0,1){90}}
\put(150,10){\line(0,1){90}} \put(155,10){\line(0,1){90}}
\put(120,100){\line(1,0){40}} \put(160,100){\line(0,-1){90}}
\put(120,100){\line(1,1){40}} \put(125,100){\line(1,1){40}}
\put(130,100){\line(1,1){18}} \put(160,130){\line(1,1){10}}
\put(135,100){\line(1,1){18}} \put(165,130){\line(1,1){10}}
\put(140,100){\line(1,1){18}} \put(170,130){\line(1,1){10}}
\put(145,100){\line(1,1){18}} \put(175,130){\line(1,1){10}}
\put(150,100){\line(1,1){18}} \put(180,130){\line(1,1){10}}
\put(155,100){\line(1,1){40}}  \put(160,100){\line(1,1){40}}
\put(120,10){\line(1,0){40}}
\put(10,100){\line(1,0){110}}\put(10,10){\line(1,0){110}}
\put(160,100){\line(1,0){110}}\put(160,10){\line(1,0){110}}
\end{picture}

\vspace*{.5cm}

 \noindent
 Looking for a separable solution of the form
\[ E_{\1}(y,z)=U(z)\,\exp(i\,n_{\I}\sin\theta\, k y)\,\,,\]
with $\theta$ representing the incidence angle, we obtain the
following second-order linear differential equation for $U(z)$,
\begin{equation}
\label{eqU} U''(z) + \left[n^{\2}(z)-n^{\2}_{\I}+
n^{\2}_{\I}\cos^{\2}\theta\,\right]\,k^{\2}U(z)=0\,\,.
\end{equation}
This equation is formally identical to the one-dimensional
Schr\"odinger equation when the factor which multiplies $U(z)$ is
replaced by $2m(E-V)/\hbar^{^{2}}$. As for Schr\"odinger the
solutions are oscillatory (travelling waves) or evanescent
(tunneling) according to whether the term in square brackets is
positive or negative respectively. A particular plane wave
solution of Eq.(\ref{eqU}), corresponding to an incoming wave in
region I, is
\[
\begin{array}{rcl}
z<0\, & \,\,\,:\,\,\, & \exp(\,i\,n_{\I}\cos{\theta}\,kz) + R\,
\exp(-\,i\,n_{\I}\cos{\theta}\,kz)\,\, ,\\ \\ 0<z<L &
\,\,\,:\,\,\, & F\,\exp\left(\,i\,n_{\I}\sqrt{ \tilde{n}^{\2} -
\sin^{\2}\theta}\,\,kz\right) +
G\,\exp\left(-\,i\,n_{\I}\sqrt{\tilde{n}^{\2} -
\sin^{\2}\theta}\,\,kz\right)\,\, ,\\  \\
%(2) &\mbox{If}\,\,\, W_{\0}=\sqrt{2}K \,:\, & A\,M_c
%\displaystyle{\left[\left(\frac{K L}{n \pi } \right)^{\2},\frac{n
%\pi}{L} \, X \right] + B\,M_s \left[\left(\frac{K L}{n \pi }
%\right)^{\2},\frac{n \pi}{L} \, X \right]} \\
z>L & : & T\,\exp(\,i\,n_{\I}\cos{\theta}\,kz)\,\,,
\end{array}
\]
with $\tilde{n}=n_{\II}/n_{\I}$ and $R$, $F$, $G$ and $T$
determined by the boundary conditions. The boundary conditions can
be set as in quantum mechanics by noting that for piece-wise
discontinuities in the ``potential''  $n(z)$ the function $U(z)$
and its first derivative $U'(z)$ must be continuous at the
boundaries, or, equivalently, since the magnetic field is
proportional to the derivative of the electric field, by the
continuity of these fields across the boundaries. Note that the
exponentials in region II are oscillatory  when $\tilde{n}^{\2} >
\sin^{\2}\theta$ and evanescent for $\tilde{n}^{\2} <
\sin^{\2}\theta$. Thus, if $\tilde{n}>1$ the solutions will always
yield a propagation wave in region II of the $y$-$z$ plane with
direction given by Snell's law. When $\tilde{n}<1$ both types of
solutions exist, with tunneling occurring when
$\sin\theta>\tilde{n}$, i.e. for incident angles greater then a
critical value $\theta_{c}$ ($\sin\theta_{c}=\tilde{n}$).

The comparison with non-relativistic quantum mechanics also
implies that the $y$-dependence replaces the time dependence of
the latter.  This accounts for the constant term in front of
$U(z)$ in Eq.(\ref{eqU}). Another identity is the condition (which
we prove in the next section) that
\[ |R|^{^2}+\,|T|^{^2}=1\,\,.\]
In quantum mechanics this implies conservation of
probability\cite{COHEN}  while here it implies conservation of
energy\cite{SKP}.

\section*{II. REFLECTION AND TRANSMISSION COEFFICIENTS}

Consider first diffusion ($\tilde{n}>\sin\theta$) and treat the
continuity conditions for $U(z)$ and $U'(z)$ at the two interfaces
($z=0,L$) independently. Let $r_{\0}$ and $t_{\0}$ be the
coefficients at the $z=0$ interface, then
\begin{eqnarray}
\label{eqa} r_{\0} & = & \left(\cos \theta -
\sqrt{\tilde{n}^{\2}-\sin^{\2}\theta}\,\right)/\left(\cos \theta +
\sqrt{\tilde{n}^{\2}-\sin^{\2}\theta}\,\right)\,\,, \nonumber \\
t_{\0} & = & 2\,\cos \theta /\left(\cos \theta +
\sqrt{\tilde{n}^{\2}-\sin^{\2}\theta}\,\right)\,\,.
\end{eqnarray}
For a wave travelling from region II to region I (e.g. a wave
reflected from the $z=L$ interface) the corresponding coefficients
$\tilde{r}_{\0}$ and $\tilde{t}_{\0}$ are
\begin{eqnarray}
\label{eqb}
\tilde{r}_{\0} & = & -\, r_{\0}\,\,, \nonumber \\
\tilde{t}_{\0} & = &
\sqrt{\tilde{n}^{\2}-\sin^{\2}\theta}\,\,t_{\0} /\cos \theta \,\,.
\end{eqnarray}
Note that for diffusion all the coefficients  in Eq.(\ref{eqa})
and (\ref{eqb}) are real. At the $z=L$ interface, we need only
consider waves impinging from the left since, for our choice of
particular solution, there is no incoming wave from the right in
region III. Thus, the only reflection and transmission
coefficients, are
\begin{eqnarray}
r_{\L} & = & \tilde{r}_{\0} \exp\left(\,2\,i\,n_{\1}\sqrt{
\tilde{n}^{\2}- \sin^{\2}\theta}\,\,kL\right) \,\,, \nonumber \\
t_{\L} & = & \tilde{t}_{\0} \exp\left[\,i\,n_{\1}\left(\sqrt{
\tilde{n}^{\2}- \sin^{\2}\theta}-\cos
\theta\,\right)\,kL\right]\,\,.
\end{eqnarray}
Now, we may calculate the $R$ and $T$ coefficients by summing
individual {\em multiple} reflection contributions, e.g. the first
contribution to $R$ will be $r_{\0}$, the second will be
$t_{\0}r_{\L}\tilde{t}_{\0}$ and so forth,
\begin{equation}
R  = r_{\0}+t_{\0}r_{\L}\tilde{t}_{\0}+
t_{\0}r_{\L}\tilde{r}_{\0}r_{\L} \tilde{t}_{\0}+ ... +
t_{\0}r_{\L}\left(\tilde{r}_{\0}r_{\L}\right)^n \tilde{t}_{\0}+
...
\end{equation}
The series converges because $0<\theta<\pi/2$ implies that
$|\tilde{r}_{\0}r_{\L}|<1$. Summing, we find
\begin{equation}
R  =  r_{\0}+t_{\0}r_{\L}\tilde{t}_{\0} /
 \left(1-\tilde{r}_{\0}r_{\L}\right)\,\,.
\end{equation}
In the same way,
\begin{equation}
 T = t_{\0}t_{\L}+
t_{\0}\tilde{r}_{\0}r_{\L} t_{\L}+ ... +
t_{\0}\left(\tilde{r}_{\0}r_{\L}\right)^n t_{\L}+ ...=
t_{\0}t_{\L} /
 \left(1-\tilde{r}_{\0}r_{\L}\right)\,\,.
\end{equation}
Consequently, using the identities $\tilde{r}_{\0}=-r_{\0}$ and
$r^{\2}_{\0}+t_{\0}\tilde{t}_{\0}=1$, we find
\begin{equation}
|R|^{^{2}} =   |r_{\0}+r_{\L}|^{^{2}} /\,
 |1+r_{\0}r_{\L}|^{^{2}}\,\,\,\,\,\mbox{and}\,\,\,\,\,
 |T|^{^{2}}  =  |t_{\0}t_{\L}|^{^{2}} /\,
  |1+r_{\0}r_{\L}|^{^{2}}\,\,.
  \end{equation}
It follows after a little algebra that
\begin{equation}
|R|^{^{2}}+\,|T|^{^{2}}=1\,\,,
\end{equation}
as anticipated. With the above expressions for $R$ and $T$, the so
called {\em wave limit} or total coherence, identical to the
quantum mechanics results, we reproduce the standard phenomena of
{\em resonance} when $|T|^{^2}=1$. This occurs when the phase in
$r_{\L}$ is such that
\[
r_{\L} = -\,r_{\0}\,\,\,\,\,\Leftrightarrow \,\,\,\,\,n_{\1}\sqrt{
\tilde{n}^{\2}- \sin^{\2}\theta}\,\,kL = n\,\pi\,\,.
\]
There is however another way to interpret the series expansions
for $R$ and $T$. Let us introduce it by simply observing a
numerical fact. If we (modulus) square the individual terms in the
series and then add, we find a different $|R|^{^{2}}$ and
$|T|^{^{2}}$,
\begin{equation}
\sum_n|R_n|^{^{2}} =
r_{\0}^{\2}+(t_{\0}r_{\0}\tilde{t}_{\0})^{^{2}} /
 \left(1-r_{\0}^{\4}\right) \neq  |R|^{^{2}} \,\,\,\,\,\mbox{and}\,\,\,\,\,
 \sum_n|T_n|^{^{2}}  =  (t_{\0}\tilde{t}_{\0})^{^{2}} /
 \left(1-r_{\0}^{\4}\right) \neq |T|^{^{2}}\,\,,
\end{equation}
with conservation of energy in the {\em particle limit}, where the
interference between individual $n$ amplitudes is null,
\begin{equation}
 \sum_n\left(|R_n|^{^{2}}+T_n|^{^{2}} \right)  =  1\,\,.
\end{equation}
These two {\em limits} are easily explained. The former {\em wave
limit} occurs when all the contribution overlap as is the case of
plane waves. The second {\em particle limit} occurs when no
overlapping occurs (see the next section). In quantum mechanics
this latter limit corresponds to wave-packets small compared to
the barrier width ($L$). The details of how the wave packets are
created is not important. It is the limit when the time taken for
a wave packet to travel back and forth in region II is sufficient
to separate the individual reflected and/or transmitted wave
packets. There are of course intermediate cases of partial
overlap. Notice that in the  particle limit there is {\em no}
resonance phenomena. Similarly, for any given  localized optical
transverse electric beam, we can calculate $|T|^{^{2}}$ for
various $L$ values (see next section) and see the transition from
typical oscillatory (resonance) shape to a constant (particle)
limit.

\section*{III. LOCALIZED BEAMS (NUMERICAL ANALYSIS)}

In quantum mechanics the particle limit is obtained by considering
narrow (compared to the barrier width) wave
packets\cite{DEL1,DEL2,DEL3,DEL4}. This is done by integrating the
plane wave results with, say, a gaussian function in particle
momentum. The optical equivalent is to integrate over the incoming
angle $\theta$ and again this can be done with a gaussian in
$\theta$ (see below). In situations in which tunneling may occur
one should formally limit the allowed values of $\theta$ to either
the diffusion or tunneling regions. In practice a strongly peaked
dependence around a mean $\theta$ value say $\theta=\theta_{\0}$
is sufficient as long as $\theta_{\0}$ is sufficiently removed
from  the critical angle $\theta_{c}$.

We shall use for our numerical calculations the following gaussian
function
\begin{equation}
 g(\alpha) =
 \frac{\sqrt{\delta}}{\left(2\pi\right)^{^{3/4}}}\,
 \exp\left[-\,\frac{\left(\alpha-\alpha_{\0}\right)^{^2}
 \delta^{^{2}}}{4}\right]\,\,,
 \end{equation}
with $\alpha=\cos \theta$ ($\alpha_{\0}=\cos \theta_{\0}$) and
$\delta=n_{\I}k\,d$.

 The integration over angles
about $\theta_{\0}$ produces a spatial localization in $z$ and $y$
(the analog of a quantum mechanics wave packet). We recall that
for our optical study all results are time independent
(stationary). The localized distributions in $z$ and $y$ are for
the incoming, reflected and transmitted beams given by
\begin{eqnarray*}
E_{\1,inc}(y,z)&=&\int_{\0}^{^{1}}\mbox{d}\alpha\,\,
g(\alpha)\,\exp(i\,\alpha\,\delta \,z_d)\,
\exp(i\,\sqrt{1-\alpha^{\2}}\,\,\delta\, y_d)\,\,,\\
E_{\1,ref}(y,z)&=& \int_{\0}^{^{1}}\mbox{d}\alpha\,\, R(\alpha)\,
g(\alpha)\,\exp(-\,i\,\alpha\,\delta\, z_d)\,
\exp(i\,\sqrt{1-\alpha^{\2}}\,\,\delta \,y_d)\,\,,\\
E_{\1,tra}(y,z)&=& \int_{\0}^{^{1}}\mbox{d}\alpha\,\, T(\alpha)\,
g(\alpha)\,\exp(i\,\alpha\,\delta\, z_d)\,
\exp(i\,\sqrt{1-\alpha^{\2}}\,\,\delta \,y_d)\,\,,
\end{eqnarray*}
with $y_d=y/d$, $z_d=z/d$ and  $L_d=L/d$.

For diffusion phenomena, and when the beam localizations are
smaller than the dimension $L$ of region II, we obtain a
zebra-like structure sketched in Fig.1(b). The various reflected
(transmitted) beams are separated in the $y$-$z$ plane. No
interference occurs between them and consequently no resonance
phenomena exists. As in quantum mechanics, these multiple
structures do not occur in tunneling phenomena\cite{HAR,REC,DEL5},
see Fig. 1(a). Indeed for tunneling the sum $\sum_n |R_n|^{^{2}}$
diverges, as does $\sum_n |T_n|^{^{2}}$, thus the individual terms
cannot be identified with physical probabilities. In tunneling
only {\em one} reflected and transmitted wave exists. The
calculation of $R_n$ and $T_n$ is then, at best, a technique for
the determination of $R$ and $T$.

We can exhibit these differences graphically by considering a
narrow
 beam incident at an angle $\theta_{\0}=\pi/4$. Two cases for $\tilde{n}$
 will be considered $\tilde{n}=1/2$ and $\tilde{n}=\sqrt{3}/2$. The
choice of a narrow beam ($\delta=50$) guarantees that the gaussian
angle distribution, centered in $\pi/4$, is practically zero for
$\theta<\pi/6$ (the critical angle for $\tilde{n}=1/2$)  and for
$\theta>\pi/3$ (the critical angle for $\tilde{n}=\sqrt{3}/2$).
Consequently, for $\tilde{n}=1/2$ we have "tunneling" and for
$\tilde{n}=\sqrt{3}/2$ we have multiple diffusion. In Fig.2 and 3,
we display, for the diffusion case, plots of $|E_{\1}|^{^{2}}$
against $z$ for fixed $y$ and against $y$ for fixed $z$. We
readily see the multiple beams. Comparison of Fig.2(a) and 2(b)
shows that the reflected beams remain separated in $z$.

\section*{IV. CONCLUSIONS}

In this paper, we have studied the behavior of a localized beam in
a stratified medium. {\em The localization is achieved by
integrating over the incidence angle}. Depending on the value of
$\tilde{n}$, we have two phenomena. One is the formation of
multiple beams, the other occurring for tunneling yields a single
reflected and transmitted beam. We have shown some examples of
these phenomena. In both cases resonance effects are absent. If we
substitute the $y$-axis with the time axis, we replicate the
results of multiple diffusion and/or tunneling in non-relativistic
quantum mechanics. There are of course some significant
difference. Foremost, the absence of $\hbar$ and the
interpretation of $|R|^{^{2}}$ and $|T|^{^{2}}$ in terms of energy
probabilities. The wave packets in quantum mechanics move in time.
In our optical model all results are stationary. This is a
significantly useful feature, since time measurements are all but
impractical in quantum mechanics\cite{REC}.

However, the analogy allows us to anticipate some further
consequences for localized optical beams. For example, while the
resonance phenomenon in tunneling is absent for a single barrier,
it surprisingly reappears\cite{DEL2} for twin or even multiple
identical barriers (always in the tunneling regime). Again this is
a consequence of interference. Thus, if the size of the barriers,
including the inter-barrier distances, is much larger than the
incoming beam, resonance will not occur and multiple beams, caused
by reflections between barriers, will appear. The ephemeral nature
of probability densities makes an optical analogy simpler to
create and study. It is even conceivable that questions related to
{\em tunneling times} for which there are diverse
definitions\cite{REC} and the related Hartman effect\cite{HAR}
could be studied experimentally. Another potential source of study
is the effect of localization on theoretical predictions almost
always based upon a plane wave analysis.

The results of this paper, with the wave and particle limits,
clearly demonstrate that these effects can be significant. In
particle physics the effect of wave packets upon phenomena such as
neutrino oscillations and oscillation phenomena in general, have
little or no possibility of experimental testing. Perhaps through
optics, we may experiment with some of these questions.

We have exhibited in our numerical analysis, and shown in our
graphs, the absence of a particle limit in the case of tunneling.
The theoretical method of calculation we have used, based upon the
sum of individual contributions still works, but only if
interpreted as an analytic continuation of the diffusion case. The
infinite series in tunneling formally diverges. This is most
simply seen by the fact in this case $r_{\0}$ is complex with {\em
unitary modulus}.

\newpage

\begin{figure}[hbp]
\hspace*{-2.5cm}
\includegraphics[width=19cm, height=24cm, angle=0]{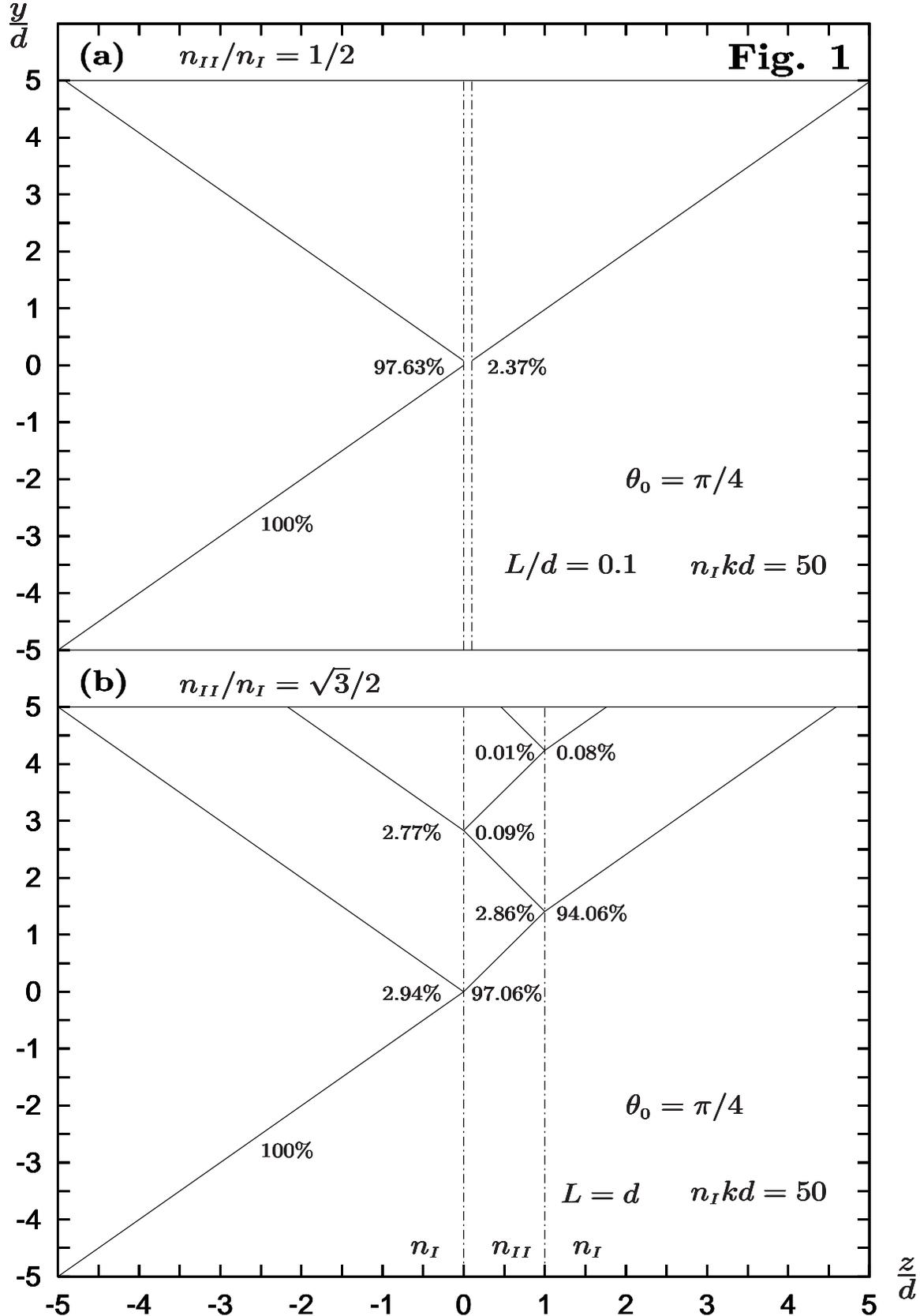}
\vspace*{-2cm}
 \caption{Tunneling (a) and diffusion (b) of a localized optical beam by a dielectric
 film. The localization is achieved by
integrating over the incidence angle (narrow gaussian distribution
around the angle $\theta_{\0}=\pi/4$).  For diffusion phenomena,
when the beam localizations are smaller than or of the order of
the dimension of region II, we obtain a zebra-like structure, i.e.
a multiple diffusion.}
\end{figure}

\newpage

\begin{figure}[hbp]
\hspace*{-2.5cm}
\includegraphics[width=19cm, height=24cm, angle=0]{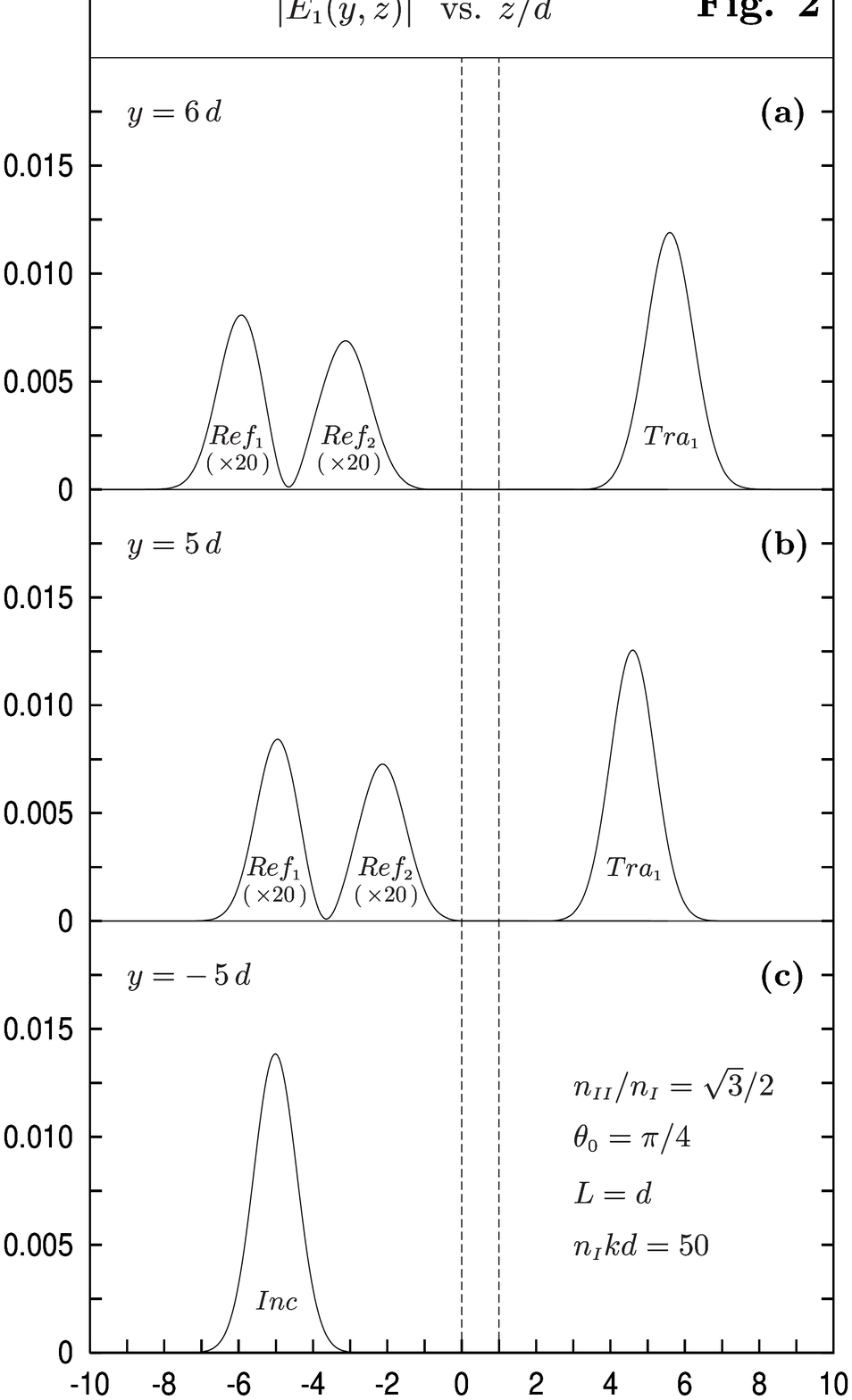}
\vspace*{-2cm}
 \caption{For the diffusion case, the plots of $|E_{\1}|^{^{2}}$ against $z$
for fixed $y$ show the multiple beams and the localization in the
$z$-axis.}
\end{figure}

\newpage

\begin{figure}[hbp]
\hspace*{-2.5cm}
\includegraphics[width=19cm, height=24cm, angle=0]{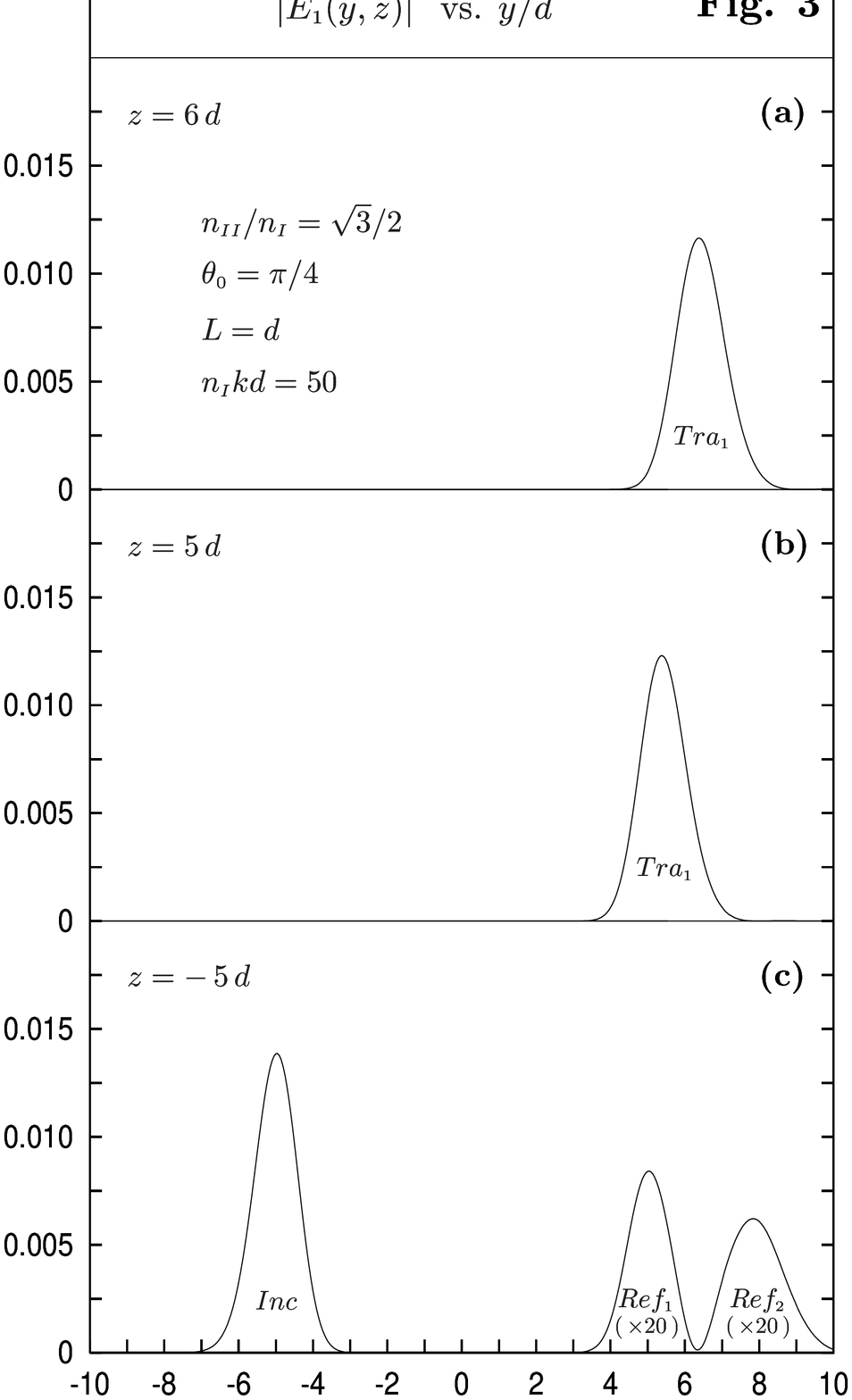}
\vspace*{-2cm}
 \caption{For the diffusion case, the plots of $|E_{\1}|^{^{2}}$ against $y$
for fixed $z$ show the multiple beams  and the localization in the
$y$-axis.}
\end{figure}


\begin{thebibliography}{99}

\bibitem{DEL1}
A. Bernardini, S. De Leo and P. Rotelli, Mod. Phys. Lett. A {\bf
19}, 2717 (2004).

\bibitem{DEL2}
S. De Leo and P. Rotelli, Phys. Lett. A {\bf 341}, 294 (2005).

\bibitem{HAR}
T. E. Hartman, J. Appl. Phys. {\bf 33}, 3427 (1962).


\bibitem{REC}
V. S. Olkhovsky, E. Recami and J. Jakiel, Phys. Rep. {\bf 398},
133 (2004).

\bibitem{DELNEU1}
S. De Leo, C. Nishi and P. Rotelli, Int. J. Mod. Phys. A {\bf 19},
677 (2004).

\bibitem{DELNEU2}
A. Bernardini and S. De Leo, Phys. Rev. D {\bf 70}, 022101 (2004).


\bibitem{BORN}
M. Born and E. Wolf, {\em Principles of optics}, Cambridge UP,
Cambridge (1999).

\bibitem{ABE}
F. Abel\'es, Ann. de Physique {\bf 5}, 596 (1950).

\bibitem{COHEN}
C. Cohen-Tannoudji, B. Diu and F. Lalo\"e, {\em Quantum
mechanics}, John Wiley \& Sons, Paris (1977).


\bibitem{SKP}
A.B. Shvartsburg, V. Kuzmiak and G. Petite, Phys. Rep. {\bf 452},
33 (2007).


\bibitem{DEL3}
S. De Leo and P. Rotelli, Eur. Phys. J. C {\bf 46}, 551 (2006).

\bibitem{DEL4}
S. De Leo and P. Rotelli, Phys. Rev. A {\bf 73}, 042107-7 (2006).



\bibitem{DEL5}
S. De Leo and P. Rotelli, Eur. Phys. J. C {\bf 51}, 241 (2007).



\end{thebibliography}
\end{document}